# A Secure Key Agreement Protocol for Dynamic Group

Muhammad Bilal and Shin-Gak Kang

*Abstract*— To accomplish secure group communication, it is essential to share a unique cryptographic key among group members. The underlying challenges to group key agreement are scalability, efficiency, and security. In a dynamic group environment, the rekeying process is more frequent; therefore, it is more crucial to design an efficient group key agreement protocol. Moreover, with the emergence of various group-based services, it is becoming common for several multicast groups to coexist in the same network. These multicast groups may have several shared users; a join or leave request by a single user can trigger regeneration of multiple group keys. Under the given circumstances the rekeying process becomes a challenging task. In this work, we propose a novel methodology for group key agreement which exploits the state vectors of group members. The state vector is a set of randomly generated nonce instances which determine the logical link between group members and which empowers the group member to generate multiple cryptographic keys independently. Using local knowledge of a secret nonce, each member can generate and share a large number of secure keys, indicating that SGRS inherently provides a considerable amount of secure subgroup multicast communication using subgroup multicasting keys derived from local state vectors. The resulting protocol is secure and efficient in terms of both communication and computation.

## I. Introduction

IN the recent past with the advent of fast networking technologies, there has been a profound increase in the speed of the Internet and the degree of connectivity. In addition, with the emergence of new Internet applications such as video conferencing, online joint workspaces, group chat, multi-user games and online social networking applications, numerous possibilities for group communications have been created. Group participants share common interests and share the responsibility of secure group communication. In group communication, agreement regarding a secure group key is one of the most important and challenging tasks. Specifically, to maintain a secure group key in a dynamic environment becomes more difficult as the reestablishment of the group key should be rapid and lightweight with regard to complexity. Secure rekeying becomes an even more challenging task in resource-limited networks such as wireless sensor networks (WSNs) [1] and body-area networks (WBAN) [2], as most conventional cryptographic mechanisms and security protocols are not suitable for resource-limited WSNs or WBANs. For example, very efficient public key algorithms, such as ECC [3], need a fraction of a second to execute encryption/decryption procedures, while a symmetric key algorithm such as RC5 [3] needs only a fraction of a millisecond to perform encryption and decryption procedures [4-5]. For computational efficiency, secure group communication it is essential, with the group key following a symmetric key algorithm.

For secure group communications, the two basic goals are authentication and confidentiality. Precisely, authentication guarantees that the communicating entity is an authorized entity, which is alive and participating in a protocol run according to a defined role. Further, the protocol run follows the correct pre-defined sequence of a protocol run, and confidentiality guarantees that the transmitted messages are recognizable and/or decrypted only by the intended entities.

In this paper, we present a unique secure group key agreement protocol suite known as SGRS (Secure Group key-agreement with Random nonce Sharing). SGRS is a decentralized and distributed protocol; it is decentralized because for scalable instances of SGRS, the groups are cascaded into larger groups (Section IV-E), and it is distributed because all of the nodes in each group contribute to computing the keying material. The SGRS protocol suite consists of four underlying protocols: the Join, Leave, Merge and Partition protocols. The common framework for these protocols is based on a unique and novel group structure, where each group member generates a secret nonce which should be known to all except one. The set of locally known nonce instances represents the state vector of a group member. In a group of $N$ members, all members have a state vector of size $|N| - 1$. Considering that the members are arranged in a logical circular linked list where a pointer to the next member indicates that the nonce of next node is unknown to the current node. For instance, $N_{i-1} \rightarrow N_i$ means that the secret nonce generated by member $N_{i-1}$ is unknown to member $N_i$; i.e., $n_{i-1} \notin S_i$ where $S_i$ is the state vector of member $N_i$, representing set of all secret nonce known to the $ith$ member, and $n_{i-1}$ is the nonce generated by member $N_{i-1}$. This novel approach of creating logical relationships among group

This work was supported by the ICT R&D program of MSIP/IITP. [R-20160302-003082, Standards development for service control and contents delivery for smart signage services].

The authors are with the University of Science and Technology Korea, Electronics and Telecommunications Research Institute (ETRI Campus), 305-700 Daejeon, Rep. of Korea (e-mail: mbilal@etri.re.kr; sgkang@etri.re.kr).

members is the key factor in our protocol. Using local knowledge of the secret nonce, each member can generate and share a large number of secure keys, indicating that SGRS inherently provides a considerable amount of secure subgroup communication using subgroup multicasting keys derived from local state vectors. We assume that an independent authentication protocol authenticates the group member. Consequently, we focus on the confidentiality aspect of secure group communication. If a member or group of members wants to join a secure communication group, they should initially be authenticated by a separate authentication protocol.

The remainder of this paper is organized as follows. In Section II, we give a brief system overview and discuss the characteristics of the group and keying system. In Section III, the proposed scheme is described in detail. Section IV presents the results of a performance analysis of SGRS against several well-known schemes. Finally, we provide concluding remarks in Section V.

## II. RELATED WORK

The group communication over IP was first presented in 1986 by Deering [6]. However, IP multicast itself does not have any mechanism to prevent group communication by non-group members. Afterward, with the emergence of new network technologies, group communication faced several new security and efficiency issues. Group communication can be secure if all group members share a common secret key, with the key generated and distributed through a secure procedure. Group communication is considered to be efficient if it has low computational and communication complexity. Various protocols have been proposed for securely and efficiently establishing a secret cryptographic key among group members. The protocols developed thus far can be categorized into three broad categories, as follows: 1) Centralized key distribution and management systems, 2) decentralized key distribution and management systems, and 3) distributed key distribution and management systems [2-5]. In the centralized approach [11,12], the overall key management complexity is low from the standpoint of a group participant; however, the centralized key management entity is associated with heavy computational and communicational complexity. The centralized approach is vulnerable to DoS/DDoS attacks and inherits the potential of a single-point failure. Moreover, tackling the issue of scalability in the centralized approach is a challenging task. In the decentralized approach [13, 14], the computational and communication complexity levels are distributed among the subgroup managers. The single-point failure is confined to the subgroup region at the expense of higher complexity of group re-keying after a join, leave, merge or partition event. In the distributed approach [15], single-point failures do not occur, but ensuring secure re-keying is a challenging task. The SGRS is a decentralized and distributed protocol, decentralized because for scalable SGRS, the groups are cascaded into larger groups (Section IV-E), and distributed because all nodes in each group contribute to computing the keying material.

In earlier work [11], the author proposed a scheme to reduce the communication complexity of the rekeying process in a dynamic group. In the proposed scheme, the head node transmits a random magic number to all group members along with a time stamp value. The member nodes undertake a left shift of the magic number and calculate the new key by an XOR operation on the old key and the left-shifted magic number. The proposed scheme requires a powerful group head node which must send $n$ number of encrypted messages for a group of size $n$, leading to the scalability problem. Seo et al. [12] proposed a group key management protocol for the cluster-based topology which establishes the group key using certificate-less asymmetric cryptography. However, this scheme is computationally expensive due to its use of asymmetric cryptography.

In work by Mehdizadeh et al. [14], the group is divided into small subgroups. The rekeying process is only confined to the locality of the event (Join and leave) in a subgroup. In the proposed scheme the network is divided into two levels: the multicast level and unicast level. The re-keying process is confined to multicast level. However, it causes an issue of data transmission within the group. The data need to be translated on each edge node of the subgroup.

In another study [16], the authors presented a group key agreement protocol suite based on blending binary key trees with the Diffie-Hellman key exchange. The usage of a hierarchal logical key significantly reduces the number of keys held by each group member. The key of each member is constructed from its child members; all participating members calculate intermediate blending keys independently and finally compute the group key.

Other work [15] presents an asymmetric group key agreement protocol, in which the general construction of the protocol is based on the well-known Chinese remainder theorem in conjunction with the NTRU cryptosystem. The overall protocol provides an efficient solution.

In another study [17], an asymmetric group key agreement protocol based on a proxy re-encryption technique was proposed. The keys are arranged in a tree format, where the members represent the key pairs and the edges represent the proxy re-encryption keys. Each participating member holds the proxy re-encryption keys from the root to the leaf member. In two of the aforementioned studies [16,17], to construct a logical tree of keys, the parent members are computed from child members; consequently, members must be synchronized, otherwise any delay can cause an interruption in a protocol run. Additionally, during the setup time, both protocols depend upon a leader; in other words, a single-point failure may still arise in the system.

In other work [18], the author presented another protocol based on the Chinese remainder theorem. This protocol constructs a tree in which each member holds two values: a key and a modulus.

Most existing group key management protocols are designed to establish a single secure group. However, with the emergence of various group-based services, it is becoming common for several multicast groups to coexist in the same network. These groups may have several common users. In some cases [13,19], the authors discussed the issue of the coexistence of multiple multicasting groups in a single network. This feature is similar to SGRS secure multicasting subgroup communication, which is also utilized in our scheme. However, one of these schemes [19] is computationally expensive because all of the keys are generated using asymmetric cryptography, which makes the overall process of key generation computationally expensive. In another scheme [13], the computational and communication cost complexity is highly dependent upon the number of multicasting groups and the number of users participating in each group. On the other hand, the establishment of the group key in SGRS is independent of the number of multicasting groups or the number of users involved in each group, as SGRS inherently empowers the group members to create secure multiple multicasting groups.

## III. SYSTEM OVERVIEW

In SGRS, the participating group members share the secret nonce and create a logical circular linked list. Using local knowledge of the shared nonce, each member can generate and share a large number of secure keys. We assume that before joining the group the requesting member/group is authenticated by an independent authentication entity using independent authentication protocol, largely discussed in literature [20-23], and further assume that the secrete nonce of sponsor member is known to the authentication entity. In the subsequent section, we explain the characteristics of the group and the keying system in detail.

**Notations**

- $N_i = ith$ member
- $n_i = $ secret nonce of $ith$ member
- $S_i = $ State Vector: Set of all secret nonce known to the $ith$ member.
- $X_s = $ Set of secret nonce created and shared among all members of a dynamic group
- $X_N = $ Set of member ids of all members in dynamic group, corresponding ids of $X_s$
- $Y_s = $ A subset of $X_s$
- $Y = $ A subset of $X_N$, which are corresponding ids of $Y_s$
- $K_Y = $ A key derived from elements of $Y_s$
- $K_P = $ Public key
- $K_G = $ Group key
- $E_A(B) = $ Encryption of B using key $K_A$, where $K_A$ is any valid $K_Y$, $K_P$ or $K_G$
- $\|$ symbol for concatenation of terms
- $\Vdash$ symbol to explicate a portion of the term
- $S \vdash m$ indicate that $S$ can generate $m$

### A. Characteristics of Group

Each group member generates a secret nonce, by using random number generator, which should be known to all except one. The local knowledge of the secret nonce represents the state vector of a group member. In a group of $N$ members, all members have a state vector of size $|N| - 1$, and each member possess a $S_i = X_s - n_{i-1}$ secret nonce. Initially, each member independently generates exactly one nonce and the state vector get updated (new nonce added or excluded) for each join, leave, merge and partition events by following the respective protocols described in subsequent Section VI.

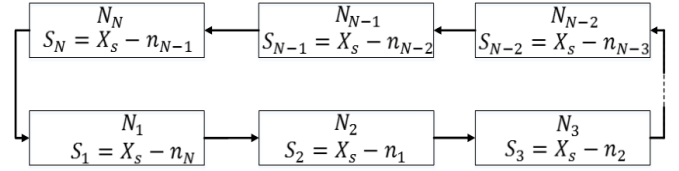

**Fig. 1.** A group of $N$ members and arranged in a logical circular linked list

For example, in Figure 1, $N_2$ knows all nonce instances expect $n_1$ and $N_3$ knows all except $n_2$, and so on. It is important to note that these members are arranged in a logical circular linked list, where a pointer to the next member indicates the secrecy of the current member's nonce to the next member. This novel approach of creating logical links among group members is the key factor in our protocol. Using local knowledge of a shared nonce, each member can generate and share a large number of secure keys.

### B. Keying System

In the SGRS scheme, each member independently generates multiple cryptographic keys $K_Y$ based upon the locally known secret set $S_i$. A multicasting key $K_Y$ is generated using the function $f(Y) = Hash(XOR(Y_s), K_G)$. To ensure the secrecy of set $Y_s$, set $Y$ is used as an index value to retrieve the valid secret nonce $Y_s$ from the state set $S_i$ of members. Only those members can generate the key $K_Y$, holding the corresponding secret nonce $Y_s$ of the enlisted member IDs in set $Y$; i.e., $K_Y$ is a valid key for all members $N_i$ if $Y_s \subseteq S_i$ and $K_Y$ is a private key of members $N_i$ if $Y_s = S_i$. This infers that the group size of $|X_N| = N$ has the following number of possible multicasting keys.

$$W = \begin{cases} \frac{N!}{(N-1)!} + \frac{N!}{(N-2)!2!} + \cdots \frac{N!}{(N-(N-2))!(N-2)!} & N \text{ is odd} \\ \frac{N!}{(N-1)!} + \frac{N!}{(N-2)!2!} + \cdots \frac{N!}{(N-(N-1))!(N-1)!} & N \text{ is even} \end{cases}$$

However, each member can participate in and/or start $Z$ number of secure subgroup communication, where $Z \subseteq W$.

$$Z = \begin{cases} \frac{(N-1)!}{(N-2)!} + \frac{N-1!}{(N-3)!2!} + \cdots \frac{(N-1)!}{(N-(N-1))!(N-1)!} + 1 & N \text{ is odd} \\ \frac{(N-1)!}{(N-2)!} + \frac{N-1!}{(N-3)!2!} + \cdots \frac{(N-1)!}{(N-(N-1))!(N-1)!} & N \text{ is even} \end{cases}$$

For instance, assuming that $N = 7$, we have a total of 119 possible keys in the system, which infers that our scheme inherits the ability to create $W = 119$ number of secure subgroups, where each member can participate in 63 subgroups.

The group key $K_G$ is generated by the hash function $K_G = Hash(n_i, X)$, where $n_i$ is the secret nonce of sponsor member $N_i$ or $N_i$ is the immediate previous member in the logical linked list if $N_{i+1}$ leaves the group and $X$ is selected based on the key update event. In join event $X$ is an existing group key, in leave and partition event $X$ is a random nonce generated and shared by the sponsor member, and in merge event it is the group key of requesting group.

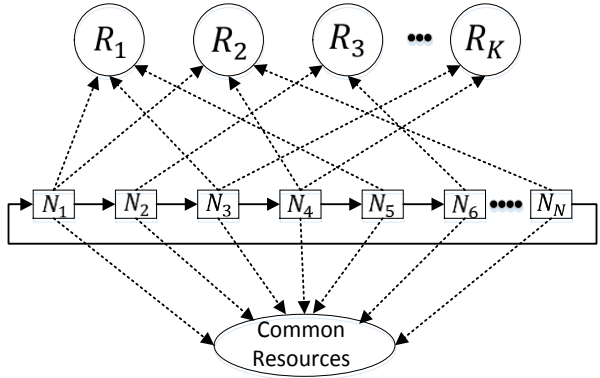

**Fig. 2.** The logically connected group members share the common resources using the group key. The subsets of group members also create multiple secure subgroups.

For computational efficiency, secure group communication it is essential, with the group key following a symmetric key algorithm. For example, very efficient public key algorithms, such as ECC [3], need a fraction of a second to execute encryption/decryption procedures, while a symmetric key algorithm such as RC5 [3] needs only a fraction of a millisecond to perform encryption and decryption procedures [4-5]. In SGRS all the cryptographic keys generated, as in the above discussion, are symmetric keys and have a size of 256 bits (32 bytes); hence, in the subsequent section of protocols any symmetric encryption supporting the 256-bit key can be used, e.g., RC5 [3]; Rijndael [24], Twofish [25], MARS [26], and Blowfish [27] symmetric encryption algorithms support the 256-bit encryption key.

## C. A Reference Framework for SGRS

The application scenario for SGRS is not limited to the group of communicating nodes. As shown in Figure 2, a group of nodes/users $\{N_1, N_2, ... N_N\}$ is arranged in a logically connected group, as explained above. All group nodes can access the common resources using the group key $K_G$ and a subset of the group can securely access resources private to the subset of nodes. Resources can be a subscription to a service, shared data, and/or a secure multicast group, for instance. Note that

As an example shown in Figure 2, $R_i \ \forall \ i = 1,2..k$, can be a service, shared memory, shared data or a multicast communication channel private to members $N_1$, $N_3$, and $N_5$. The private key can be calculated by the subgroup members independently without an exchange of a single message, as follows:

1- $Y_s = S_1 \cap S_3 \cap S_5$
2- $K_Y = f(N_1, N_3, N_5, N_7, .. N_{N-1}) = Hash(XOR(Y_s), K_G)$

Note that the function $f$ takes the list of node IDs as an input argument and translates it into vector $Y_S$ by selecting the corresponding secret nonce from the local state vector $S_i$. In the subsequent section, we demonstrate that to maintain the characteristics of a group, we share the node IDs ($Y$) and let the member node translate the vector $Y$ into $Y_s$. This ensures that only eligible nodes with the required information ($Y_s$) can generate the secret key.

At this stage as shown in Figure 3, we consider that the network is divided into clusters/cells. Each cluster provides a complete set of resources. If a user who is subscribed to multiple subgroup resources moves from one cluster to another, it will invoke a single leave and single join protocol. Once a user becomes part of a logical group, it can generate all of the subgroup keys to access the subscribed resources. As shown in Figure 3, $N_3$ in cluster 1 is subscribed to $R_1, R_2$, and $R_k$, which indicates that $N_3$ is part of one major group as well as three subgroups. When it leaves cluster 1 and joins cluster 2 as $N_4$, cluster 1 runs leave protocol while cluster 2 runs join protocol. The rekeying of three subgroups in both clusters requires a single message exchange, and all subgroup keys can be generated by the members independently using function $f$.

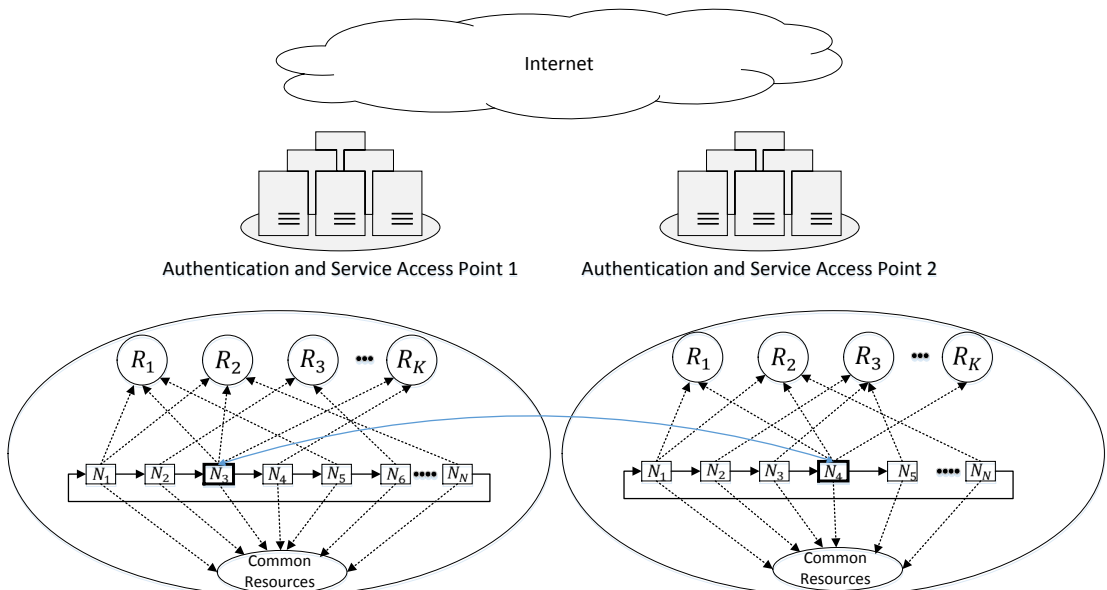

**Fig. 3.** An example when one of the group member moves from one authentication as service point to another authentication as service point. Note that whole network is divided into multiple clusters and each cluster provides a complete set of resources.

## IV. PROPOSED SCHEME

The SGRS protocol suite consists of four underlying protocols: the join, leave, merge and partition protocols. All of these protocols share a joint framework with the following prominent features:

- Each group member contributes to maintaining its state vector to preserve the characteristics of a group as defined in Section II-A.
- Upon the addition or removal of members, all members update the previous keying material based upon the updates shared by the sponsor node.
- Each member can participate and/or start $Z$ instances of secure subgroup communication, where $Z \subseteq W$.

Upon each membership event, all members independently update the state vector and compute all possible keys on demand. SGRS protocols are highly distributed; all participating members equally participate in a protocol run except for the sponsor member, which performs a few additional operations.

### A. Join Protocol

The join protocol is initiated when a potential member sends a valid join message to a sponsor member. The authentication server generates a 'joining tag' for new member $N_j$, as shown below.

$$E_G(S_j) || E_Y\left(sig(n_j)\right) \Vdash Y = \{N_i\}$$

A valid joining tag consists of two parts. The first part $E_G(S_j)$ is destined for new member $N_j$ only, whereas the second half $E_Y\left(sig(n_j)\right) \Vdash Y = \{N_i\}$ is used by new member $N_j$ to initiate the join protocol. Due to the encryption with unknown keys, the contents of the tag are hidden from the requesting member $N_j$. The first half is encrypted with the updated group key $K_G = Hash(K_G^{current}, n_i)$, where $n_i$ is the secret nonce of the current sponsor member of the group, which is the new member, who intends to join. This part of the tag consists of the vector state for member $N_j$, as the state is encrypted with the updated group key $K_G$, and the requesting node $N_j$ does not have knowledge of the contents until and unless it initiates and participates in the joining protocol. During the join protocol run, the sponsor member verifies the join request using the second half of the tag. Note that the second half of the joining tag is encrypted by multicasting key $K_Y$, which is derived from the secret nonce of the group sponsor node. This key is also unknown to the joining member. Once $N_j$ is verified, the sponsor node then shares the updated group key $K_G$ such that the joining member retrieves the state vector and becomes part of logical circular linked list of the group.

After receiving a valid join tag, $N_j$ multicasts the join request $E_Y(n_j) \Vdash Y = \{N_i\}$), and the protocol proceeds as follows:

1. $N_j \to (X_N - N_{i+1}) : E_Y\left(sig(n_j)\right) \Vdash Y = \{N_i\}$
2. $\forall (X_N - N_{i+1}) \vdash K_G^{new} = Hash(K_G^{current}, n_i)$
3. $N_i \to N_j : E_Y(K_G^{new}) \Vdash Y = \{N_j\}$
4. $N_{i-1} \to N_{i+1} : E_G^{current}(n_i)$
5. $N_j \vdash K_G^{new} = Hash(K_G^{current}, n_i)$

**1)** In first step $N_j$ multicasts the join message encrypted with multicasting key $K_Y$ derived from $N_i$'s secret nonce, which infers that all group members excluding $N_{i+1}$ receive $n_j$. All members confirm the authenticity of the sender by verifying the signatures of the authentication server and sender. **2)** All group members, excluding $N_j$ and $N_{i+1}$, generate a new group key and update their state vectors by adding the secret of the new member $N_j$. **3)** The sponsor node $N_i$ shares the new group key with the joining member $N_j$. At this point, the newly joining member $N_j$ can decrypt the first half of the join ticket to acquire its state vector $S_j$. **4)** In parallel with message 3, the member $N_{i-1}$ shares the nonce of the sponsor node with member $N_{i+1}$. This sharing breaks the logical link between the sponsor member and $N_{i+1}$ and established a new logical connection between $N_j$ and $N_{i+1}$. **5)** Finally, after receiving $n_j$, the member $N_{i+1}$ also generates a group key.

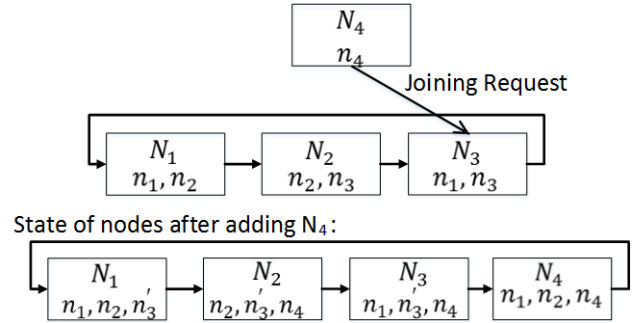

Fig. 4. Update of state vectors in a simple scenario when a new member joined the group

We consider a group of three members, as shown in Figure 4, where $N_3$ is the sponsor member and $N_4$ sends a valid join request $E_Y\left(sig(n_4)\right) \Vdash Y = \{N_3\}$, which infers that only $N_2$ and $N_3$ can decrypt the message using $K_Y = f(\{N_3\})$. For the given example the join protocol proceeds as follows: **1)** $N_2$ and $N_3$ verify the request and add $n_4$ to the state vector. **2)** $N_2$ and $N_3$ generate a new group key $K_G^{new} = Hash(K_G^{current}, n_3)$ and update their state vectors by adding the secret of the new member $N_4$. **3)** The sponsor node $N_3$ shares the new group key with joining node $N_4$. **4)** In parallel with message 3, the member $N_2$ shares the nonce of the sponsor node with member $N_1$. This sharing breaks the logical link between $N_3$ and $N_1$ and establishes a new logical link between $N_4$ and $N_1$. **5)** Finally, after receiving $n_3$, member $N_1$ also generates group key $K_G = Hash(K_G^{current}, n_3)$.

*B. Leave Protocol*

The leave protocol is initiated when a valid member becomes invalid, for instance, when the group subscription time expires or if a valid member unsubscribes from the group membership. In either case, the sponsor member of the group is responsible for initiating the leave protocol. Here, $N_j$ is the sponsor of the group, and $N_i$ is the departing member. Upon the occurrence of a leave event, the sponsor $N_i$ initiates the leave protocol, which proceeds as follows:

1. $N_j \rightarrow X'_N : E_Y(n_{random}) \,||\, E_G^{old}(Y) \Vdash Y = \{N_{i-1}\}$
2. $\forall (X'_N - N_{i+1}) \vdash (n_{i-1}, K_G) = \left(Hash(n_{i-1}^{old}, n_i), Hash\left(Hash(n_j^{old}, n_{random}), n_{random}\right)\right)$
3. $N_j \rightarrow N_{j+1} : E_Y(K_G) \,||\, E_G^{old}(Y_N) \Vdash Y = \{N_{i-1}^{old}\}$

**1)** When $N_i$ leaves the group, the sponsor member $N_j$ updates its secret nonce $n_j = Hash(n_j^{old}, n_{random})$ and generates a new group key $K_G = Hash(n_j, n_{random})$. The sponsor member also updates its state vector by updating the secret nonce $n_{i-1} = Hash(n_{i-1}, n_i)$ of member $N_{i-1}$ but keeps the old value as well until the second step is completed. The sponsoring member skips this step if the sponsor is next to the leaving member $N_i$ in the logical linked list, i.e., if $j = i + 1$. After generating the new group key and updating the state vector, the sponsor member multicasts the partition message in conjunction with $n_{random}$. The multicasting key $K_Y$ is derived from set $Y = \{N_{i-1}\}$, which infers that only remaining members $X'_N$ will receive this message. Note that even if the sponsor node updated $n_{i-1}$, it will still use the old value of $n_{i-1}$ for the multicast message, until the second step completed. The departing node $N_i$ cannot decrypt this message as $n_{i-1} \notin S_i$, hence departing node $N_i$ cannot generate the new group key; it ensures the forward secrecy. **2)** After receiving the arguments for generating a new group key, all members excluding $N_{i+1}$ generate a new group key and update the state vectors by updating nonce $n_{i-1}$ by hashing its old value with $n_i$. This step ensures that departing member cannot generate the group key, as $n_{random}$ is unknown to the departing member. Moreover, it ensures that member $N_{i+1}$ should not generate $n_{i-1}$, as $n_i$ is unknown to $N_{i+1}$. Hence, after the second step, $n_{i-1} \notin S_{i+1}$, meaning that a new logical link between $N_{i-1}$ and $N_{i+1}$ is created. 3) Finally, the sponsor member shares the group key with $N_{i+1}$.

State of nodes before $N_4$ left the group:

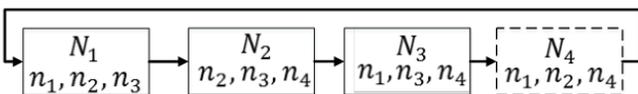

State of nodes after $N_4$ removed from the group:

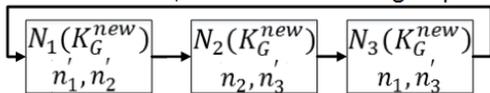

Fig. 5. Update of state vectors in a simple scenario when a group member leave the group

Let us consider a group of four members, as shown in Figure 5, where $N_2$ is the sponsor member, and $N_4$ is the departing member. For the given example the leave protocol proceeds as follows: **1)** When $N_4$ leaves the group, the sponsor node $N_2$ updates its secret nonce $n'_2 = Hash(n_2, n_{random})$, generates a new group key $K_G = Hash(n'_2, n_{random})$, and multicast the partition message in conjunction with $n_{random}$. The multicasting key $K_Y = f(\{N_3\})$ is derived from set $Y_s = \{N_3\}$, which means $N_4$ cannot receive the message. The sponsor member also updates its state vector by updating the secret nonce $n'_3 = Hash(n_3, n_4)$. **2)** After receiving the arguments for generating new group key, $N_1$ updates the state vector by updating the secret nonce $n'_2 = Hash(n_2^{old}, n_{random})$ and generates new group key $K_G = Hash(n'_2, n_{random})$. Besides, $N_3$ updates the state vector by updating the secret nonce $n'_3 = Hash(n_3, n_4)$. At this stage, $N_4$ (departing node) cannot generate $K_G$ and $n_3$ as $n_{random}$ is unknown to the $N_4$. **3**-Finally, $N_2$ shares the updated group key using the key derived from old $n_3$. The old $n_3$ is unknown to $N_4$, hence the group key is only delivered to $N_1$ and $N_3$. The member $N_1$ updates its state vector by deleting old $n_3$.

*C. Merge Protocol*

To merge $k$ number of small groups into one large group, the merge protocol runs concurrently in $\lceil \log_2 k \rceil$ rounds. For each run of the protocol, we obtain a merged group of two subgroups. For example, as shown in Figure 6, six small groups $\{G_1, G_2, G_3, G_4, G_5, G_6\}$ are merged into one large group $G_{16}$ in $\lceil \log_2 6 \rceil = 3$ rounds.

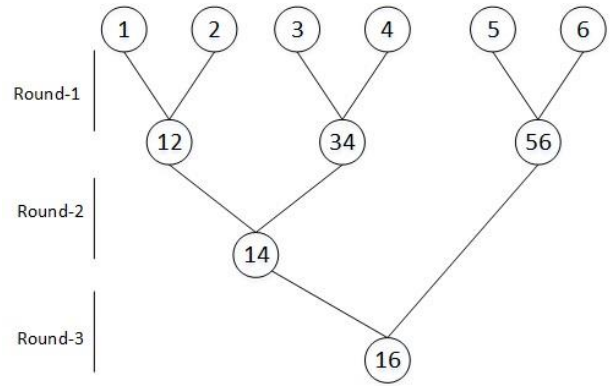

Fig. 6. An example of merging 6 small groups into a single larger group. The merge completed in $\lceil \log_2 6 \rceil = 3$ rounds.

The merge protocol is initiated when a sponsor member sends a valid join message to a sponsor member of another group. Here, $N_i^a$ is the current sponsor of the group $a$ and $N_i^b$ is the sponsor of group $b$. After receiving a valid join ticket $(E_Y(K_G^b || sig(X_s^b)) || n_{i-1}^a \Vdash Y = \{X_N^a - N_{i+1}^a\})$, the members of group $b$ know the secret of the immediate previous member in the logical linked list of $N_i^a$. Upon sending the join request, the protocol proceeds as follows:

1. $N_i^b \rightarrow N_i^a : E_Y(K_G^b || sig(X_s^b)) \Vdash Y = \{X_N^a - N_{i+1}^a\}$
2. $N_i^a \rightarrow (X_N^a - N_{i+1}^a) : E_G^a(Update)$
3. $N_i^a \rightarrow (X_N^b - \{N_1^b\}) : E_Y^b(S_i^a) \,||\, E_G^b(Y) \Vdash Y^b = \{N_2^b\}$

4. $N_i^a \to N_1^b: E_Y(n_m^b, (S_i^a - n_i^a))||E_G^b(Y^b) \Vdash Y^b = \{N_2\}$
5. $N_i^a \to (X_N^a - N_{i+1}^a) : E_G^a(E_Y(X_S^b||K_G^b)||Y^a) \Vdash Y^a = \{N_i\}$
6. $N_{i-1}^a \to N_{i+1}^a: E_G^a(E_Y(S_{i-1}^a, (X_S^b - n_m^b))||K_G^b)||Y^a) \Vdash Y^a = \{N_{i+1}^a\}$

**1)** The sponsor of group $b$ sends the join message, encrypted with key $K_Y$, as derived from $\{X_N^a - N_{i+1}^a\}$, inferring that only the sponsor members of group $a$ receive the Join message. Sponsor member $N_i^a$ confirms the authenticity of the sender by verifying the signatures. **2)** The sponsor of group $a$ multicasts the updated message, meaning that all group members excluding $N_{i+1}$ update the value of $n_i^a = Hash(n_i^a, K_G^a)$. $N_{i+1}$ cannot update because $n_i^a \notin S_{i+1}$. The updated value of $n_i^a$ ensures backward secrecy. **3)** $N_i^a$ shares his state vector with all members of group $b$, except the tail member $N_1^b$. To ensure this, $N_i^a$ multicasts the message using $K_Y$, which is derived from the secret of head member $N_m^b$ i.e., $n_m^b$. **4)** $N_i^a$ shares a set $\{n_m^b, (S_i^a - n_i^a)\}$ of secret values with $N_1^b$ and straight away, member $N_1^b$ breaks the logical link with $N_m^b$ and establishes a new logical link with member $N_i^a$. Note that messages 2 and 3 are sent by the sponsor member of group $a$, but they are encrypted with keys derived from group $b$. The rationale behind this choice is to prevent the disclosure of $S_i^a$ from other members of group $a$. **5)** At this stage, the sponsor member shares the vector $X_S^b$ and $K_G^b$ with all members of group $a$, excluding $N_{i+1}^a$. The message is encrypted with $K_G^a$ and further encrypted with $K_Y$ derived from $N_i^a$; this infers that the message is prevented from being accessed by $N_{i+1}^a$ and all members of group $b$. **6)** $N_{i-1}^a$ shares a set $\{n_i^a, (X_S^b - n_m^b)\}$ of secret values with $N_{i+1}^a$, and straight away member $N_{i+1}^b$ breaks the logical link with $N_i^b$ and establishes a new logical link with member $N_m^b$, with group $b$ added between $N_i^a$ and $N_{i+1}^a$. All members of groups $b$ and $a$ shared the common group key $K_G^b$ and updated the state vectors with new secret values while maintaining the logical links.

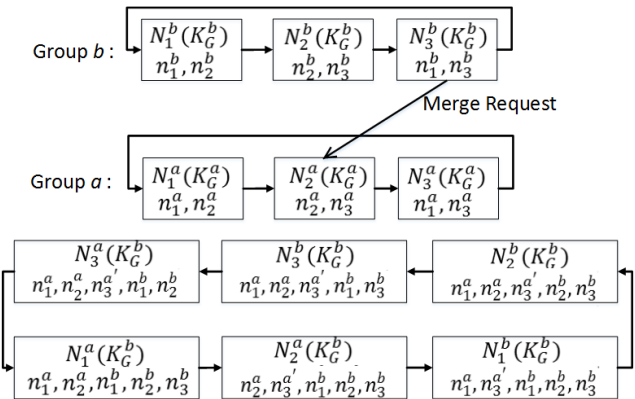

**Fig. 7.** Update of state vectors in a simple scenario when two groups merged and share a common group key

Let's consider two groups $a$ and $b$ each with three member nodes, as shown in Figure 7, where $N_3^b$ sponsor member of groups $b$ sends a valid merge request to $N_2^a$, the sponsor member of group $a$. For the given example the merge protocol proceeds as follows:. **1)** $N_2^a$ verifies the request, and updates his secret nonce $n_2^{a'} = Hash(n_2^a, K_G^a)$ and state vector $S_2^a$. **2)** $N_2^a$ informs all group members, excluding $N_3^a$ to updated state vectors by updating $n_2^{a'} = Hash(n_2^a, K_G^a)$. **3)** $N_2^a$ shares his state vector with the members of group $b$, except tail member $N_1^b$, to guarantee this, $N_2^a$ multicast the message using $K_Y$ which is derived from the secret of head member $N_3^b$ i-e, $n_3^b$. **4)** $N_2^a$ shares a set $\{n_3^b, (S_2^a - n_2^a)\}$ of secret values with $N_1^b$; this ensures that the member $N_1^b$ broke the logical link with $N_3^b$ and established a new logical link with member $N_2^a$. **5)** Now sponsor member $N_2^a$ shares the vector $X_S^b$ and $K_G^b$ with all members of the group $a$ excluding $N_3^a$. The message is encrypted with $K_G^a$ and further encrypted with $K_Y$ derived from $n_2^a$; this infers the message is prevented from being accessed by $N_3^a$. **6)** $N_1^a$ shares a set $\{n_2^a, (X_S^b - n_3^b)\}$ of secret values with $N_3^a$, which means the member $N_3^a$ broke the logical link with $N_2^a$ and established a new logical link with member $N_3^b$, in other words, group $b$ is added between $N_2^a$ and $N_3^a$.

### D. Partition Protocol

Assume that there is a group of $X_N$ members where $X_M$ members leave the group simultaneously and are left with $X_N'$ members. Occasionally, a fault in the network disconnects a large number of nodes simultaneous. Upon the occurrence of a partition event, the current sponsor of group $N_i$ will initiate the partition protocol, which proceeds as follows:

1. $N_i \to X_N': E_Y(n_{random}) ||E_G^{old}(Y_N) \Vdash Y = \{\cap_j S_j, \forall N_j \in X_N'\}$
2. $\forall(X_N' - N_{i-1}) \vdash S_j = G(S_j) \wedge K_G = Hash(n_i, n_{random})$
3. $N_i \to N_{i-1}: E_Y(K_G) ||E_G^{old}(Y) \Vdash Y_N = \{\cap_j S_j, \forall N_j \in X_N'\}$

**1)** Upon the occurrence of a partition event, the sponsor member $N_j$ updates its secret nonce $n_j = Hash(n_j^{old}, n_{random})$ and generates a new group key $K_G = Hash(n_j, n_{random})$. The sponsor member multicast the partition message in conjunction with $n_{random}$. The multicasting key $K_Y$ is derived from the set $Y_N = \{\cap_j S_j, \forall N_j \in X_N'\}$, which infers that only remaining members $X_N'$ will receive this message. This also infers that all members $n_i \in X_N'$ remove all members $n_k \in X_M$ from the local secret list. **2)** After having received the arguments for generating a new group key, all members excluding $N_{i-1}$ generate the new group key and update the state vectors by running function $G$. Function $G$ updates the state vector of all remaining members by discarding all of the secrets which cannot be generated and those which belong to $X_M$. For instance, if $\{N_{i+1}, ... N_j\} \in X_M$, then function $G$ updates $n_i = Hash(n_i, n_j)$, where $n_j$ is the secret of the head member among contiguous departing members in the logical linked list. In the worst case, when leaving members are noncontiguous, function $G$ will perform $|X_N - X_M|$ number of hash operations to update each member's state vector. In best case, when all leaving members are contiguous, function $G$ must perform only one hash operation to update each member's state vector. **3)** Sponsor member $N_i$ sends the group key to $N_{i-1}$ explicitly, as node $N_{i-1}$ does not know the value of $n_j$.

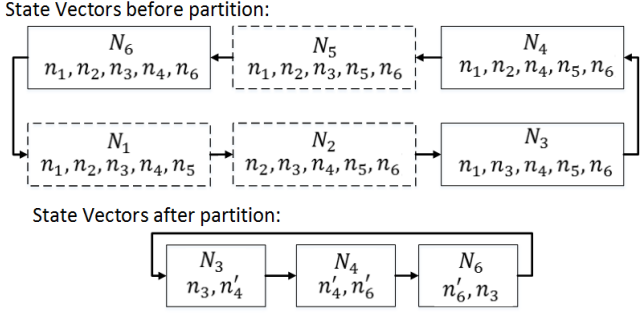

**Fig. 8.** Update of state vectors in a simple scenario when a group is partitioned

We consider a group of six members $\{N_1, N_2, N_3, N_4, N_5\} = X_6$, as shown in Figure 8, where $N_3$ is the sponsor member and $\{N_1, N_2, N_5\}$ leave the group simultaneously. For the given example the partition protocol proceeds as follows: **1)** The sponsor member $N_3$ generates a new group key $K_G = Hash(n_3, n_{random})$ and multicast the partition message, including the new group key and a list of members $X_M = \{N_1, N_2, N_5\}$, to $N_4$ and $N_6$ using encryption key $K_Y$ derived from $Y_S = \{n_1, n_4, n_6\}$. **2)** After receiving the partition message, all remaining members update the state vector, where $n'_3 = Hash(n_3, n_5)$   $n'_4 = Hash(n_4, n_5)$ and $n'_6 = Hash(n_2, n_6)$ and generate the group key $K_G = Hash(n_3, n_{random})$.

### E. Scalable SGRS for Larger Group

For each join and leave message, we are required to perform a hash operations proportional to the group size. SGRS may encounter a scalability problem when the group size reaches millions of users. For instance, in a group of one million members, each member must maintain a state vector of 999999 nonce instances. In terms of memory consumption, this will require approximately 40MB, which is not an issue for modern end-user devices, but for each leave and join event, the protocol requires the performance of nearly one million hash operations. With such a large group size, the probability of the occurrence of a leave or join event also increases.

The problem of scalability can be solved by dividing the large group into smaller cascaded groups arranged in multiple layers, as shown in Figure 9. Let us consider $N$ number of members divided into $k$ number of groups with each group $(G_i)$ having its group key $(K_G^i \ \forall \ i = 1,2,3..k)$ shared and generated, as discussed in relation to the SGRS protocol suite. All of these groups can generate a supergroup $(SG_i)$ by considering each group $(G_i)$ as a logical member of the supergroup and considering the corresponding group key $K_G^i$ as their secret nonce. With this type of arrangement, all of the groups and group members share a common group key generated at the supergroup level. The supergroup $(SG_i)$ can further be cascaded to generate an ultra-supergroup $(USG_i)$, and so on.

The cascaded group limits the required number of nonce updates for the new group key to the lowest group level. For example, consider two levels which are cascading, where groups join to create a supergroup $(SG)$. If a leave or join event occurs in group $G_i \in SG$, the SGRS will run locally and will produce the new group key $K_G^i$, which will further serve as an updated nonce of group member $G_i$ at the supergroup level. At the supergroup level, all of the group members then generate the new group key, as follows:

1. $SGRS(G_i) \vdash K_G^i$ (New local group key)
2. $\forall \ (G_j \in SG - G_{i-1}) \vdash K_{SG} = Hash(K_{SG}, K_G^i)$
3. $Any \ G_j \to G_{i-1}: E_Y(K_G) \Vdash Y_N = \{G_{i+1}\}$

**1)** An SGRS event occurs, leading to production of new local group key $K_G^i$. **2)** At the supergroup level, all group members, excluding $G_{i-1}$, generate new a supergroup key and share it with local members. **3)** Any other group members $G_j$ can send the updated group key to $G_{i+1}$.

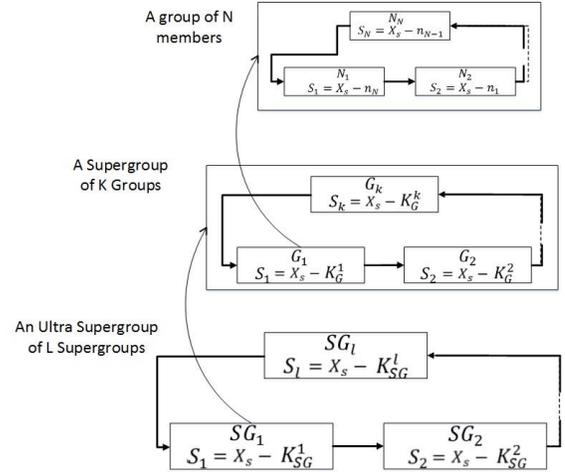

**Fig. 9.** A cascaded group example , $k$ number of groups generate one super-group and $l$ number of super-groups generate one large ultra-super-group. The group key generated at ultra-super-group level is shared among all super-group, groups, and group members.

For a cascaded group solution for the scalable SGRS scheme, we make one important assumption. At the group level, the sponsor member is considered to be a permanent and trusted member who does not share upper-level keys, except for the final supergroup key, with local members.

## V. PERFORMANCE ANALYSIS

### A. Security Analysis

There are four major security apprehensions regarding group communication: group key secrecy, backward secrecy, forward secrecy and key independence [3,6]. SGRS addresses all of these security concerns.

1- Group key secrecy: In SGRS, it is computationally infeasible to generate a group key unless the intruder knows the secret state vector of the group members.

TABLE I
A COMPUTATION AND COMMUNICATION COST COMPARISON OF GROUP KEY PROTOCOLS

| Schemes | PROTOCOL | Comp. Complexity | Comm. Complexity | Comm. Cost in bytes |
|---|---|---|---|---|
| Kim et. al [16] | Join | $(3 \log_2 N + 9)Ex + (4 + 2N)E$ | $2BC$ | $N\;CK + N\;CK\;(2N-1)$ |
| | Leave | $(3 \log_2 N + 9)Ex + (2 + 2N)E$ | $1BC$ | $N\;CK\;log_2 N$ |
| | Merge | $(3 \log_2 N + 9)Ex + (2k(1+N) + 1 + N)E$ | $(1 + 2k)BC$ | $\frac{CK}{k}(kN - N)(2N - k) + N(2n-1)\;CK$ |
| | Partition | $(3 \log_2 N + 9)Ex + miN(2k, N/2)E$ | $min(2k, N/2)BC$ | $k\;N\;CK\;log_2 N$ |
| X. Lv et. al [15] | Join | $(2N^2 + N)Mu + (5 + N)E$ | $2UC + 1BC$ | $(2N + 1)Int + N\;CK$ |
| | Leave | $(2N^2 + N)Mu + (1 + N)E$ | $1BC$ | $N\;Int$ |
| | Merge | $(2N^3 + N2 + N)Mu + (5 + N)E$ | $kBC + 2kUC$ | $(N^2 - \frac{N^2}{K})(CK - Int)$ |
| | Partition | $(2|Gi|2 + |Gi|)Mu + (5 + N)E$ | $kBC$ | $(N^2 - \frac{N^2}{K})Int$ |
| Chen [17] | Join | $(4 \log_2 N)Ex + (5 + 2N \log_2 N)E$ | $(2log2N)MC + 2BC$ | $CK(2 \log_2 N + 2) + N\;CK(\log_2 N + 1)$ |
| | Leave | $(2 \log_2 N)Ex + 2h\;(2 + 2N \log_2 N)E$ | $(2log2N)MC$ | $CK2log2N + N\;CK(\log_2 N + 1)$ |
| Mehdizadeh et al. [14] | Join | $(6N^2 + 3N)Mu + (8 + 4N)E$ | $3UC + 2MC$ | $2int + CK\left(K + \frac{N}{K} + 1 + \log_2 K\right)$ |
| | Leave | $(6N^2 + 3N)Mu + (6 + 4N)E$ | $4UC + 2MC$ | $4int + CK\left(K + \frac{N}{K} - 3 + \log_2 K\right)$ |
| Zhong et al. [13] | Join | $\left(8 + \frac{N}{K} + 2^{K+1} + \log_2 K\right)E + 2Ex$ | $1UC + 3BC$ | $5NCK + 1Int$ |
| | Leave | $\left(9 + \frac{N}{K} + 2^{K+1} + \log_2 K\right)E + Ex$ | $3BC$ | $5NCK$ |
| SGRS | Join | $(N - 1)H + (N + 2)E$ | $2UC + 1BC$ | $N\;Int + CK$ |
| | Leave | $2NH + 2NE$ | $2UC + 1BC$ | $(N - 1)Int + CK$ |
| | Merge | $(2K - 1)\left[\left(7 + \frac{N}{K}\right) + \left(\frac{N}{K} - 1\right)H\right]$ | $(3K - 3)UC + (3K - 3)BC$ | $4KCK + \left(3 + \frac{N}{K}\right)\frac{N}{K}Int$ |
| | Partition | $(N + 2K)E + (K + N - 2)H$ | $KBC + KUC$ | $NInt + K\;CK$ |

Please note the following for Table-1

$E$=encryptions/decryptions; $Ex$= modular exponentiations; $Mu$ = multiplications; $H$= hash operation; $K$= number of groups to be merged, partitioned or coexists in same network; $|Gi|$ = size of ith group; $UC$= Unicast Message, $BC$= Broadcast Message, $k$=Number of Groups, $CK$= 32 bytes (Represents Size of Cryptographic Key, Hash, and Signature), $Int$= 4 bytes( Represents Size of the nonce, Node Ids and Integers).

2- Backward secrecy: To ensure backward secrecy, when a member or group of members joins the group, they are prevented from regenerating the previous group keys. As the group keys are generated using shared secrets, at the time when a member or group of members joins the group, the secret nonce of the sponsor member has been updated using a one-way hash function. Thus, upon the occurrence of a join event, the requisite material to re-generate the previous group key is destroyed. In the Join and leave protocol, this step ensures this property.

3- Forward secrecy: When a member or group of members leaves the group, they are precluded from knowing/generating future group keys. In both algorithm 2 and algorithm 4, step 2 ensures this property. As the group keys are generated using shared secrets, at the time when a member or group of members leaves the group, we update the secret state vectors of all group members using a one-way hash function. Note that upon a leave event, it is not necessary to update the entire state vector, and the multicasting keys are generated by including a new group key as an argument which ensures that multicasting keys cannot be used by leaving members.

4- Key Independence: In SGRS, group key generation is independent of previously generated group keys. Any knowledge of previously known group keys cannot help discover any other group key.

### B. Efficiency Analysis

This section analyzes the efficiency of SGRS in terms of computational and message complexity against earlier works [13-17]. We consider that encryption should protect all types of keying materials exchanged among group members. Further, we assume that encryption and decryption have identical computational costs, represented by $E$; for instance, the exchange of an encrypted unicast message increases the computational cost by $2E$, whereas an encrypted broadcast message increases the computational cost by $(1 + N)E$. Similarly, the encrypted multicast message adds $(1 + k)E$ to the total computational cost. Here, $K$ denotes the size of the multicasting group. Additionally, we presume that the computational price of an asymmetric key pair and an asymmetric proxy re-encryption key are identical.

To present the computational analysis of the results of earlier work [13,14], we make a few extra assumptions. In Mehdizadeh et al. [14], the group is divided into two network levels: the multicast level and the unicast level. We assume that the network of size $N$ is divided into $K$ subgroups, each of size $|G_i| = N/k$. Hence, in Mehdizadeh et al. [14], we have $2^{k+1} - 1$ nodes at the multicast level and $N$ number of nodes at the unicast level. In the performance analysis section, the author overlooked the rekeying cost induced at the multicast level. At the multicast level, the keys are arranged in a logical key hierarchy (LKH) and the key tree is a binary balanced tree. Upon each key update event, the multicast server must send at least one multicast message to all leave $2^{k+1} - 1$ nodes. In Zhong et al. [13] the authors discussed the issue of the coexistence of multiple multicasting groups in a single network. However, in that work [13], the computational and communication cost complexity is highly dependent on the number of multicasting groups and the number of users participating in each group. If there are $K$ number of multicasting services, in upper case, we have $2^k - 1$ key encryption keys (KEKs) while in lower case, if all users are subscribed to one multicast service, there will simply be a single KEK. For our analysis, we consider the lower case scenario.

Table I presents a computational cost comparison of SGRS and the approaches by Kim et al. [16], Lv et al. [15], Chen [17] Mehdizadeh et al. [14] and Zhong et al. [13]. All of the proposed schemes require at least $N$ number of encryption/decryption operations because there is at least one broadcast message in all types of protocol runs. However, in two works ([16] and [17]) the significant contribution of the computational cost is made by the exponential operations, requiring $O(logN)$ modular exponential operations. In two other approaches [14,15], multiplication operations are the major contributors, and both require $O(N^2)$ multiplication operations. Zhong et al. [13] is very efficient, but that scheme requires the re-encryption and translation of each message by the subgroup leader, whereas in SGRS, the significant contribution to the computational cost stems from the one-way hash function, requiring $O(N)$ hash operations for the join and leave protocols, while it is limited to the number of groups and the size of the group in the merge and partition protocols. In SGRS, the computational workload is well distributed among the group members. For instance, in the join protocol, each member performs one hash operation, while in the leave protocol, each member performs two hash operations. SGRS is persistent, and during the protocol run, the availability of the correct group key is independent of the failure of one or a set of members.

Table I also presents the communication complexity and a cost comparison of SGRS and the approaches by Kim et al. [16], Lv et al. [15], Chen [17] Mehdizadeh et al. [14] and Zhong et al. [13]. The communication complexity determines the number of messages exchanged, whereas communication cost determines the total amount of data exchanged during the protocol run. In terms of the communication complexity and considering the number of join and leave protocols, the approaches by Zhong et al. and by Chen [13, 17] are the most expensive. Considering join and leave protocol, SGRS and the approaches by Lv et al. and Kim et al. [15] and [16] have very low and almost similar levels of communication complexity. For the merge protocol, SGRS is expensive compared to two earlier studies [15] and [16], while in the case of a partition protocol, the approach by Kim et al. [16] is the most expensive protocol.

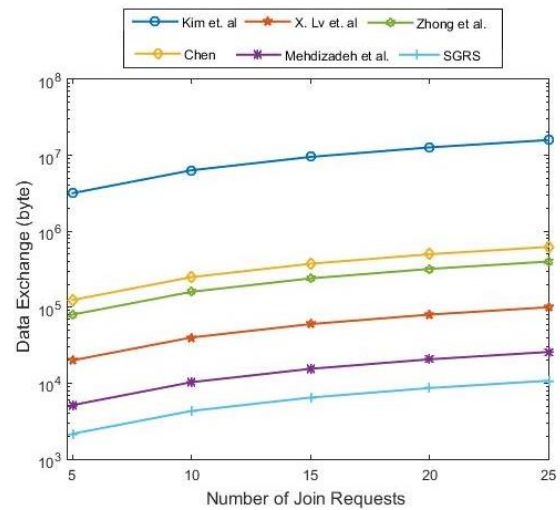

Fig. 10. The total amount of data exchanged for different number of join requests with group size of 100

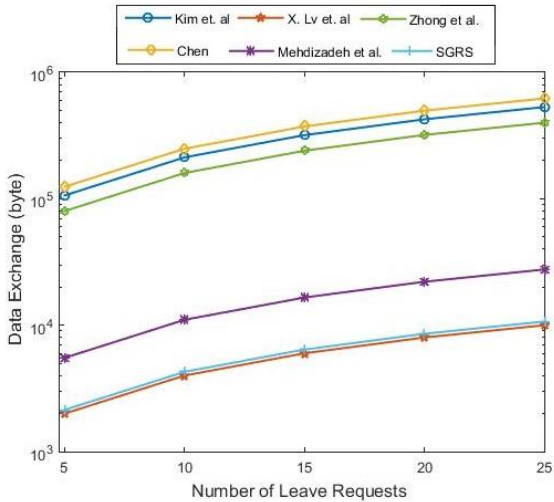

**Fig. 11.** The total amount of data exchanged for different number of leave requests with group size of 100

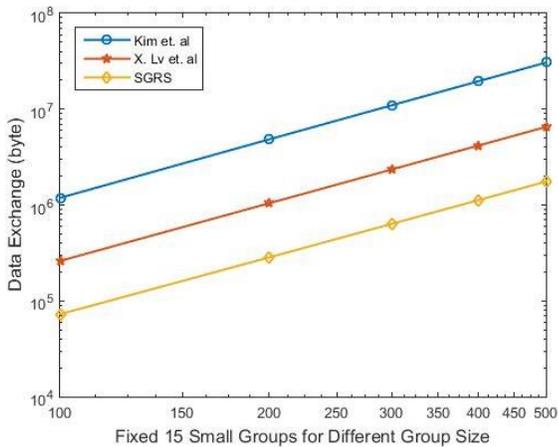

**Fig. 12** The total amount of data exchanged for different numbers of merge events assuming that the size of each group is 15, combined to create a larger group 100-50 in size

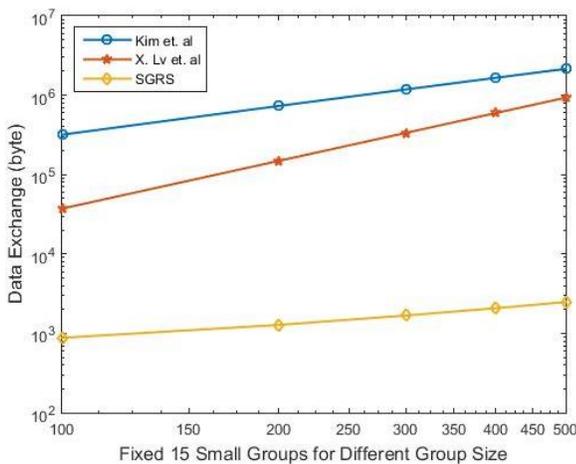

**Fig. 13.** The total amount of data exchanged for different number of partition events assuming size of each partitioned group is 15

However, the communication cost concerning the total data exchange gives a different conclusion. For simplicity of determining the communication, during merge and partition protocols, we consider that the merging or partitioning groups are of identical sizes. Further, we consider the best case condition for Kim et al. [16]; for instance, we assume that a blinded keys tree is always a perfect binary tree with at most $2N-1$ nodes. In a merge protocol after tree sharing, we consider that the protocol require one BC to share the group key, while in the partition protocol, we consider in each round sponsor node must broadcast one blinded key. To calculate the total amount of data exchanged, we assume that the integers, nonce, and the member IDs are 4 bytes in size. The cryptographic keys, hash, and signatures size are considered to be 32 bytes. The communication cost results are shown in Figure 10-13.

Figure 10 shows the comparison results of the total amount of data exchanged for the join protocol assuming a group of 100 members receiving 5 to 25 joining requests. It is quite clear that SGRS outperform all of the other schemes. Figure 11 depicts the comparison results of the total amount of data exchanged for a leave protocol considering a group of 100 members receiving 5 to 25 leave requests. It is quite clear that SGRS and the scheme by X. Lv et al. [15] outperform the remaining schemes. SGRS is slightly expensive compared to that by X. Lv et al. [15], but the performance gap is too small. Figures 12 and 13 depict the comparison results considering the total amount of data exchanged for the merge and partition protocols, respectively. In the merge protocol, we consider that 15 identically sized groups are merged to create one single large group, while we assume that one single large group is partitioned into 15 small identically sized groups; the small group size varies from 7 to 34. It is quite clear that SGRS outperform both Kim et al. [16] and X. Lv et al. [15].

## VI. CONCLUSION

In this paper, we have proposed a novel group key agreement scheme for the dynamic group, SGRS. Our scheme is distributed yet does not require synchronization among group members to share and update the keys. However, in the case of cascaded membership events, group members should perform all necessary update operations, especially updating the local group key, before moving to the next membership event. Additionally, our solution inherently provides a considerable amount of secure subgroup multicast communication using subgroup multicasting keys derived from state vectors. Moreover, SGRS establishes a symmetric group key, which ensures that group communication is computationally efficient.

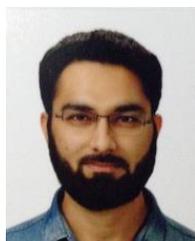

**Muhammad Bilal** has received his BS degree in computer systems engineering from University of Engineering and Technology, Peshawar, Pakistan and MS in computer engineering from Chosun University, Gwangju, Rep. of Korea. Currently, he is Ph.D. student at University of Science and Technology, Korea at Electronics and Telecommunication Research Institute Campus, Daejeon, Rep. of Korea. He has served as a reviewer of various international Journals including IEEE Communications Letters. He has also served as a program committee member on many international. His main research interests are Design and Analysis of Network Protocols, Network Architecture, and Future Internet.

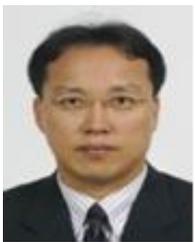

**Shin-Gak Kang** received his BS and MS degree in electronics engineering from Chungnam National University, Rep. of Korea, in 1984 and 1987, respectively and his Ph.D. degree in information communication engineering from Chungnam National University, Rep of Korea in 1998. Since 1984, he is working with Electronics and Telecommunications Research Institute, Daejeon, Rep. of Korea, where he is a principal researcher of infrastructure standard research section. From 2008 he is a professor at the Department of Information and Communication Network Technology, University of Science and Technology, Korea. He is actively participating in various international standard bodies as a Vice-chairman of ITU-T SG11, Convenor of JTC 1/SC 6/WG 7, etc. His research interests include multimedia communications and Applications, ICT converged services, contents networking, and Future Network.